\documentclass[conference,letterpaper]{IEEEtran}
\IEEEoverridecommandlockouts

\usepackage[dvipsnames]{xcolor}
\usepackage[utf8]{inputenc} 
\usepackage{comment}
\usepackage[T1]{fontenc}
\usepackage{url}
\usepackage{ifthen}
\usepackage{cite}
\usepackage{amsmath}
\usepackage{mathtools,cuted}
\usepackage{amssymb} 
\usepackage{cite}
\usepackage{pgf,tikz,pgfplots}
\usepackage{algorithmic}
\usepackage{graphicx}
\usepackage{tabularx}
\usepackage{multirow}
\usepackage{stackengine}                            

\usepackage{tikz}
\usepackage{ctable}
\usepackage{stfloats}

\newcommand{\otoprule}{\midrule[\heavyrulewidth]}

\newcommand{\RomanNumeralCaps}[1]
    {\MakeUppercase{\romannumeral #1}}
\newcommand{\que}{\mathord{?}} 
\makeatletter
\renewcommand\env@matrix[1][\arraystretch]{%
  \edef\arraystretch{#1}%
  \hskip -\arraycolsep
  \let\@ifnextchar\new@ifnextchar
  \array{\c@MaxMatrixCols c}}
\makeatother

\newcommand{\bm}{\boldsymbol{m}}
\newcommand{\by}{\boldsymbol{y}}
\newcommand{\bx}{\boldsymbol{x}}
\newcommand{\bv}{\boldsymbol{v}}
\newcommand{\bz}{\boldsymbol{z}}
\newcommand{\dv}{d_\mathsf{v}}
\newcommand{\dc}{d_\mathsf{c}}
\newcommand{\mL}{\mathcal{L}}
\newcommand{\mH}{\mathcal{H}}
\newcommand{\mZ}{\mathcal{Z}}
\newcommand{\xhat}{\hat{x}}
\newcommand{\bxhat}{\hat{\boldsymbol{x}}}
\newcommand{\Xhat}{\hat{X}}
\newcommand{\Xtilde}{\tilde{X}}
\newcommand{\balpha}{\boldsymbol{\alpha}}
\newcommand{\bbeta}{\boldsymbol{\beta}}
\newcommand{\Ma}{\mathcal{M}_{\alpha}}
\newcommand{\Mb}{\mathcal{M}_{\beta}}
\newcommand{\ma}{\boldsymbol{m}_{\alpha}}
\newcommand{\mb}{\boldsymbol{m}_{\beta}}

\newcommand{\bs}[1]{\ensuremath{\boldsymbol{#1}}}

\begin{document}
\title{On the Universality of Spatially Coupled LDPC Codes Over Intersymbol Interference Channels} 

\author{
\IEEEauthorblockN{Mgeni Makambi Mashauri\IEEEauthorrefmark{1}, Alexandre Graell i Amat\IEEEauthorrefmark{2}, and Michael Lentmaier\IEEEauthorrefmark{1} }
\IEEEauthorblockA{\IEEEauthorrefmark{1}Department of Electrical and Information Technology, Lund University, Lund, Sweden}
\IEEEauthorblockA{\IEEEauthorrefmark{2}Department of Signals and Systems, Chalmers University of Technology, Gothenburg, Sweden}\\\vspace*{-1cm}

\thanks{This work was supported in part by the Swedish Research Council (VR) under grant \#2017-04370. The simulations were performed on resources provided by the Swedish National Infrastructure for Computing (SNIC) at center for scientific and technical computing at Lund University (LUNARC).}
}
\maketitle

\begin{abstract}
In this paper, we derive the exact input/output transfer functions of the optimal a-posteriori probability channel detector for a general ISI channel with erasures. Considering three   channel impulse responses of different memory as an example, we compute the BP and MAP thresholds for regular spatially coupled LDPC codes with joint iterative detection and decoding. When we compare the results with the thresholds of ISI channels with Gaussian noise we observe an apparent inconsistency, i.e., a channel which performs better with erasures performs worse with AWGN. We show that this anomaly can be resolved by looking at the thresholds from an entropy perspective. We finally show that with spatial coupling we can achieve the symmetric information rates of different ISI channels using the same code.    
\end{abstract}

\section{Introduction}

Spatial coupling is a powerful concept that improves the belief propagation (BP) decoding threshold of the coupled system to the maximum a-posteriori (MAP) decoding threshold of the underlying uncoupled system. Spatial coupling was initially introduced in the context of low-density parity-check (LDPC) codes \cite{Fel99,Len10} and subsequently applied to other classes of codes \cite{Smi12,Mol17} and scenarios beyond the realm of coding \cite{Kud2010}.

Threshold saturation was first proven for spatially-coupled LDPC (SC-LDPC) codes for transmission over the binary erasure channel \cite{Kud11} and later for the general class of binary memoryless symmetric channels \cite{Kud13}.  Threshold saturation of SC-LDPC codes for transmission over channels with memory was  addressed in \cite{Ngu2012} for intersymbol-interference (ISI) channels with general noise model. Particularly, the authors derived  the BP generalized extrinsic information transfer (GEXIT) curves of the corresponding uncoupled systems, from which the MAP threshold can be estimated. The computation of the exact GEXIT curves requires knowledge of the input/output transfer functions of the a-posteriori probability channel detector, which are in general not available in closed form. Thus, one needs to resort to Monte
Carlo methods to provide an estimate. For the particular case of the dicode channel with erasures \textemdash known also as the  dicode erasure channel (DEC)\textemdash, in \cite{10.5555/959778} Pfister derived the corresponding transfer function. This was used in \cite{Ngu2012} to obtain the exact density evolution equations for SC-LDPC codes over the DEC, which are needed for the computation of the GEXIT curves. The authors then showed numerically  that threshold saturation occurs for the DEC.



In this paper, we consider SC-LDPC codes for general ISI channels. Our main contribution is the derivation of the transfer functions  for general ISI channels with erasures, which allows us to derive the exact density evolution equations for these channels. We then use these equations to compute BP and MAP thresholds for SC-LDPC codes over these channels. The numerical results show that, for large enough coupling memory, threshold saturation occurs for all considered channels. Furthermore, by increasing the Tanner graph density, we show that the BP thresholds approach the symmetric information rates (SIRs) of the corresponding channels, supporting the conjecture in \cite{Ngu2012} that SC-LDPC codes can universally approach the SIR of ISI channels. We further consider SC-LDPC codes over ISI channels with additive white Gaussian noise (AWGN), which reveal the same behavior.

\section{System Model}

\begin{figure}[t!]
\centerline{\includegraphics[scale=0.31]{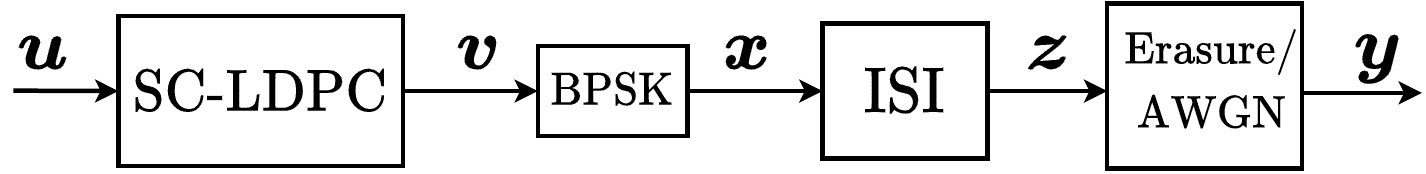}}
\caption{Block diagram showing the transmitter and the ISI channel. }
\label{model1}
\vspace{-2.5ex}
\end{figure}

The system model under consideration is shown in Fig.~\ref{model1}. A binary information sequence  $\boldsymbol{u}$ is first encoded by an SC-LDPC code onto codeword $\boldsymbol{v}$. We consider binary transmission,  with mapping $0\mapsto +1$ and $1\mapsto -1$, over an ISI channel with either erasures or AWGN. In both cases, the modulated sequence $\bx$ is first transmitted over an ISI channel. The sequence at the output of the ISI filter is denoted by $\bz$, whose elements take values on a finite alphabet $\mZ$. For the ISI-erasure channel, each element of $\bz$ is erased with probability $\varepsilon$. In this case, the elements of the received sequence, $\boldsymbol{y}$, take values on the finite alphabet $\{\mZ\,\cup\, ?\}$, with symbol $?$ denoting an erasure. For the ISI-AWGN channel, $\bz$ is corrupted by AWGN, and the elements of $\boldsymbol{y}$ take on real values.
\begin{figure}[t!]
\centerline{\includegraphics[scale=0.18]{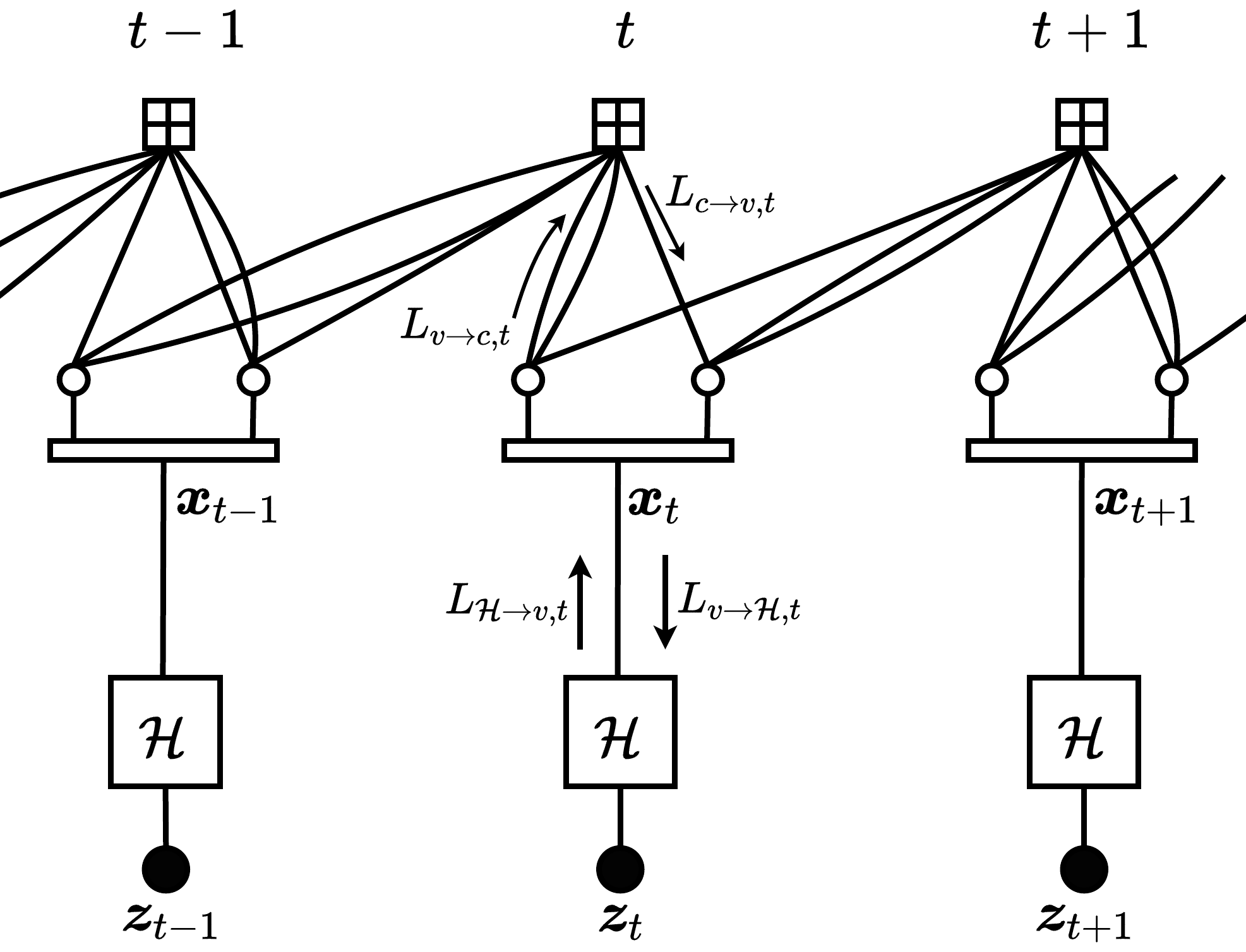}}
\vspace{-2ex}
\caption{Compact graph representation for equalization with a (3,6) SC-LDPC code with coupling memory $m=1$.}
\label{messages1}
\vspace{-2.5ex}
\end{figure}

At the receiver, we consider joint iterative decoding and channel detection, usually referred to as turbo equalization. In particular, we consider BP decoding of the SC-LDPC code and optimal channel detection, performed using the BCJR algorithm over the corresponding trellis. Graphically, decoding is performed over the factor graph depicted in Fig.~\ref{messages1}, which combines the Tanner graph of the SC-LDPC code (upper part) and the trellis of the ISI channel (lower part). Specifically, the factor graph is constructed by placing $L$ copies of a $(\dv,\dc)$ regular LDPC code of variable node (VN) degree $\dv$ and check node (CN) degree $\dc$ in $L$ spatial positions in the range $\mL\in\{1,\ldots,L\}$.  Fig.~\ref{messages1} shows the factor graph for three spatial positions, $t-1$, $t$, and $t+1$.  Each spatial position consists of $N$ VNs, represented by empty circles, and $M$ CNs, represented by squares with a cross. The $L$ copies are coupled as follows: each VN at position $t\in\mL$ is connected to CNs in the range $[t,\ldots,t+m]$, where $m$ is referred to as the coupling memory. Hence, each CN at position $t$ is connected to VNs in the range $[t-m,\ldots,t]$. The trellises of the ISI channel at each spatial position are  represented by a square labeled with the letter $\mH$, referred to as factor node. 
The VNs represented by the black circles at the bottom of the figure correspond to the symbol sequences at the output of the ISI channel (prior to the addition of the erasures or AWGN), denoted by $\{\bz_t\}$. Note that  $\bz=(\bz_1,\ldots,\bz_L)$ (see Fig.~\ref{model1}). The rectangles at each spatial position between the Tanner graph of the SC-LDPC code and the channel factor nodes represent multiplexers that multiplex the $N$ code bits (VNs) at each spatial position into a single binary sequence ($\bx_t$, with $\bx=(\bx_1,\ldots,\bx_L)$) at the input of the channel.
Decoding is then performed by iteratively passing messages along the edges of the graph in Fig.~\ref{messages1}.



\section{Input/Output Transfer Function of the BCJR Detector for an ISI Channel with Erasures}
\label{sec:transferfunction}

 In this section, we derive the transfer functions of the BCJR channel detector  for arbitrary ISI channels with erasures. These transfer functions characterize the output extrinsic erasure probabilities\textemdash from the channel detector to the SC-LDPC decoder\textemdash as a function of the a-priori erasure probabilities and the channel erasure probability. Particularly, we follow  a similar approach to that in \cite{1258535,Mol17} for  binary convolutional codes over the binary erasure channel, which  scales well with the channel memory. Compared to the case of binary convolutional codes, however, the nonbinary alphabet of general ISI channels and the lack of symmetry, which precludes assuming that the all-zero codeword is transmitted, makes the derivation for   a bit more complex. 
 
 We consider an ISI channel with memory $\nu$  and trellis states $s_1,s_2,\ldots,s_{2^\nu}$. Let $\by$ and $\bv$ be the received vector (affected by erasures) and the output of the ISI filter, respectively. Note that for ISI channels with erasures the messages exchanged between the decoder and the channel detector take values on the ternary alphabet $\{+1,-1,?\}$. We denote by $\bxhat$ the message vector from the decoder to the channel detector.

 We define the forward and backward state metric vectors at time $\tau$, $\tau=1,\dots,n$, where $n$ is the length of the ISI channel trellis,  
 as $\boldsymbol{\alpha}_{\tau}=(\alpha_{\tau}(s_1),\ldots,\alpha_{\tau}(s_{2^\nu}))$ and $\boldsymbol{\beta}_{\tau}=(\beta_{\tau}(s_1),\ldots,\beta_{\tau}(s_{2^\nu}))$, respectively. Note that $\balpha_\tau$ and $\bbeta_\tau$ are probability vectors. In the case of erasures, the vectors $\balpha_\tau$ and $\bbeta_\tau$ take values on a finite set. The sets of values that vectors $\boldsymbol{\alpha}_{\tau}$ and
 $\boldsymbol{\beta}_{\tau}$ can take on are denoted by $\Ma=\{\ma^{(1)},\ldots,\ma^{(|\Ma|)}\}$ and $\Mb=\{\mb^{(1)},\ldots,\mb^{(|\Mb|)}\}$, respectively, of cardinality $|\Ma|$ and $|\Mb|$. For the particular case of the DEC, which has memory one and the impulse response given in Table~\ref{Tab:channels} (CH-I), there are two possible states, $s_1=+1$ and $s_2=-1$, and  $\mathcal{M}_\alpha=\mathcal{M}_\beta=\left\{(1,0),(0,1),(0.5,0.5)\right\}$. Note that, in general, $\mathcal{M}_\alpha$ and $\mathcal{M}_\beta$ may differ.
 
 The sequences $\dots,\boldsymbol{\alpha}_{\tau-1},\boldsymbol{\alpha}_{\tau},\boldsymbol{\alpha}_{\tau+1},\dots$ and $\dots,\boldsymbol{\beta}_{\tau+1},\boldsymbol{\beta}_{\tau},$ $\boldsymbol{\beta}_{\tau-1},\dots$ form each a Markov chain, which can be properly described by a probability transition matrix, denoted by $\boldsymbol{M}_{\alpha}$ and $\boldsymbol{M}_{\beta}$, respectively, where the element $(i,j)$ of $\boldsymbol{M}_a$, $i\in\{1,\ldots, |\boldsymbol{M}_a|\}$, is the probability of transition from state
$\bm_a^{(i)}$ to state $\bm_a^{(j)}$, with $a\in\{\alpha,\beta\}$. 
 
 
 Our aim is to derive $\boldsymbol{M}_{\alpha}$ and $\boldsymbol{M}_{\beta}$ for an arbitrary ISI channel with erasures. We provide some details for the forward recursion. Let $X\in\{+1,-1\}$, $Z\in\mZ$, and $Y\in\{\mZ\,\cup\,?\}$  be the random variable corresponding to the symbol at the input of the ISI filter, the output of the ISI filter, and the received symbol for a given trellis section. Further, let $\Xhat\in\{+1,-1,?\}$ be the random variable corresponding to the incoming message from the decoder. The corresponding realizations are $x$, $z$, $y$, and $\xhat$. We denote by $\varepsilon$ the channel erasure probability, i.e., the probability that $Y$ is an erasure, and by $\delta$ the average probability of erasure of a message from the decoder to the channel detector, i.e., the probability that $\Xhat$ is an erasure.

 Define $P(\boldsymbol{m}_{\alpha}^{(j)}|\boldsymbol{m}_\alpha^{(i)})$ as the probability of transition from state $\bm_{\alpha}^{(j)}$ to state $\bm_{\alpha}^{(i)}$ in the forward recursion.
\begin{table}[!t]
\caption{Discrete impulse responses of the considered  ISI channels}
\vspace{-3.5ex}
\begin{center}
\begin{tabular}{cll}
	\toprule
	CH-\RomanNumeralCaps{1} & $\bs{h}=[\begin{array}{cc} 0.7071 & -0.7071 \end{array}]$ &  $\nu=1 $\\[0.1cm]
	CH-\RomanNumeralCaps{2} & $\bs{h}= [\begin{array}{ccc} 0.408 & 0.816&0.408 \end{array}]$ &  $\nu=2$\\[0.1cm]
	 CH-\RomanNumeralCaps{3} & $\bs{h}=[\begin{array}{ccccc} 0.227 & 0.46&0.688&0.46&0.227 \end{array}]$ &  $\nu=4 $\\[0.05cm]
	 \hline
\end{tabular}
\end{center}
\label{Tab:channels}
\vspace{-5ex}
\end{table}
This probability can be written as 
\begin{align}\label{pxgst}
P(\boldsymbol{m}_{\alpha}^{(j)}|\boldsymbol{m}_\alpha^{(i)})&=\sum_{\mathclap{\xhat,y}}P(\xhat,y,\boldsymbol{m}_{\alpha}^{(j)}|\boldsymbol{m}_\alpha^{(i)})\nonumber\\
  &=\sum_{\mathclap{\xhat,y}}P(\xhat|\boldsymbol{m}_\alpha^{(i)})P(y,\boldsymbol{m}_{\alpha}^{(j)}|\xhat,\boldsymbol{m}_\alpha^{(i)})\,, 
\end{align}
where we used the shorthand notation $P(\xhat,y)=P(\Xhat=\xhat,Y=y)$.

Now let $\mathcal{Z}_{ij}$ be the set of all possible values that $Z$ can take in the transition between state $\boldsymbol{m}_\alpha^{(i)}$ and state $\boldsymbol{m}_{\alpha}^{(j)}$. Then we have
\begin{align}
 P(y,\boldsymbol{m}_{\alpha}^{(j)}|\xhat,\boldsymbol{m}_{\alpha}^{(i)}) &=\sum_{z\in \mathcal{Z}_{ij}}P(y,\boldsymbol{m}_{\alpha}^{(j)},z|\xhat,\boldsymbol{m}_{\alpha}^{(i)})\nonumber\\
 &=\sum_{z\in \mathcal{Z}_{ij}}P(z,\boldsymbol{m}_{\alpha}^{(j)}|\xhat,\boldsymbol{m}_\alpha^{(i)})P(y|z) \ .
 \end{align}

This follows from the fact that $Y$ is independent of $\{\Xhat,\alpha_t,\alpha_{t+1}\}$ given $Z$. Clearly,
$P(Y=?|z)=\varepsilon$ and $P(Y=z|z)=1-\varepsilon$.

To compute $P(z,\boldsymbol{m}_{\alpha}^{(j)}|x,\boldsymbol{m}_\alpha^{(i)})$, we need to  consider two cases. If $\Xhat$ is an erasure, we have $P(z,\boldsymbol{m}_{\alpha}^{(j)}|\xhat,\boldsymbol{m}_\alpha^{(i)})=P(z,\boldsymbol{m}_{\alpha}^{(j)}|\boldsymbol{m}_\alpha^{(i)})$ for all $z\in \mathcal{Z}_{ij}$, since $Z$ is no longer constrained by $\Xhat$. On the other hand, if $\Xhat$ is not an erasure,  we get $P(z,\boldsymbol{m}_{\alpha}^{(j)}|\xhat,\boldsymbol{m}_\alpha^{(i)})=P(z,\boldsymbol{m}_{\alpha}^{(j)}|x,\boldsymbol{m}_\alpha^{(i)})$, where $P(z,\boldsymbol{m}_{\alpha}^{(j)}|x,\boldsymbol{m}_\alpha^{(i)})$ can be zero for some combinations of $\{ z,x,\boldsymbol{m}_\alpha^{(i)}\}$. In summary, we obtain
\begin{align} 
P(\xhat,y,\boldsymbol{m}_{\alpha}^{(j)}|\boldsymbol{m}_\alpha^{(i)})=  \left\{
\begin{array}{ll}
      \frac{1}{2}\bar{\delta}\bar{\varepsilon}P(z,\boldsymbol{m}_{\alpha}^{(j)}|x,\boldsymbol{m}_\alpha^{(i)}) & x,z \vspace{0.12cm} \\
     \frac{1}{2}\bar{\delta}\varepsilon & x,\que \vspace{0.12cm}\\
     \delta\bar{\varepsilon}P(z,\boldsymbol{m}_{\alpha}^{(j)}|\boldsymbol{m}_\alpha^{(i)}) &\que,z \vspace{0.12cm} \\
     \delta\varepsilon & \que,\que
\end{array}\,,
 \right.
 \label{eq:eq3}
 \end{align}
where we used the shorthand notation $\bar\delta\triangleq 1-\delta$ and $\bar\varepsilon\triangleq 1-\varepsilon$.
\begin{figure*}[!t]
\begin{equation}\nonumber
  \scalebox{0.835}{$
\frac{\varepsilon^5\left(- \delta^6\varepsilon^6(\varepsilon^{4} + 4\varepsilon^3 - 6\varepsilon^2 + 4\varepsilon - 1) + \delta^5\varepsilon^4(2\varepsilon^5 - 8\varepsilon^4 + 12\varepsilon^3 - 4\varepsilon^2 - 6\varepsilon + 4) - \delta^4\varepsilon^3(2\varepsilon^4 + 5\varepsilon^3 - 14\varepsilon^2 + 25\varepsilon - 1) + \delta^3\varepsilon^2(8\varepsilon^4 - 18\varepsilon^3 + 20\varepsilon^2 - 18\varepsilon + 8) + \delta^2\varepsilon(8\varepsilon^3 - 20\varepsilon^2 + 4\varepsilon + 8) + 40\delta\varepsilon^2 - 40\delta\varepsilon + 64\right)}{(\delta^3\varepsilon^7 - 2\delta^3\varepsilon^6 + 2\delta^3\varepsilon^5 - 2\delta^3\varepsilon^4 + \delta^3\varepsilon^3 + 2\delta^2\varepsilon^3 - 2\delta^2\varepsilon^2 + 4\delta\varepsilon^4 - 4\delta\varepsilon^3 + 4\delta\varepsilon^2 - 4\delta\varepsilon + 8)^2}
$}  
\end{equation} 
\hrulefill
\vspace{-3.5ex}
\end{figure*}
%
\begin{table}[!t]
\caption{Possible observations for DEC channel CH-I with $\alpha_t \text{$=$}\boldsymbol{m}_\alpha^1$}
\vspace{-3.5ex}
\begin{center}
\begin{tabular}{cccccccc}
	\toprule
$\Xhat,Y$& $0$,$0$ & $0$, $\que$& $\que$, $0$& $1$, $-2$ & $1$,  $\que $& $\que$, $-2$& $\que$, $\que$\\[0.1cm]
$\alpha_{t+1}$ & $\boldsymbol{m}_\alpha^1$ & $\boldsymbol{m}_\alpha^1$ & $\boldsymbol{m}_\alpha^1$ & $\boldsymbol{m}_\alpha^2$ & $\boldsymbol{m}_\alpha^2$ & $\boldsymbol{m}_\alpha^2$ & $\boldsymbol{m}_\alpha^3$ \\[0.1cm]

Prob&  $\frac{1}{2}\bar{\delta}\bar{\varepsilon} $& $\frac{1}{2}\bar{\delta}\varepsilon$& $\frac{1}{2}\delta\bar{\varepsilon}$&$\frac{1}{2}\bar{\delta}\bar{\varepsilon}$& $\frac{1}{2}\bar{\delta}\bar{\varepsilon} $& $\frac{1}{2}\bar{\delta}\varepsilon$& $\delta\varepsilon$\\[0.1cm]
\bottomrule
\end{tabular}
\end{center}
\label{Tab:tabledicode}
\vspace{-3.5ex}
\end{table}

In Table~\ref{Tab:tabledicode}, we give the  probabilities $P(\xhat,y,\boldsymbol{m}_{\alpha}^{(j)}|\boldsymbol{m}_\alpha^{(1)})$ in~\eqref{eq:eq3} for the DEC. For this channel, whose impulse response is given in Table~\ref{Tab:channels} (CH-I), $ \mZ=\{-2,-1,0,+1,+2\}$ (when no normalization by $1/\sqrt{2}$ is applied to the channel taps). The complete probability transition matrix $\boldsymbol{M}_\alpha$ is given by 
\renewcommand{\arraystretch}{1.2}
\begin{equation*}
\boldsymbol{M}_\alpha=\left[
\begin{array}{c c c}
\frac{1}{2}(1-\delta\varepsilon) & \frac{1}{2}(1-\delta \varepsilon) & \delta \varepsilon \\
\frac{1}{2}(1-\delta\varepsilon) & \frac{1}{2}(1-\delta\varepsilon) & \delta\varepsilon \\
\frac{1}{2}-\frac{\delta}{4}(1+\varepsilon) & \frac{1}{2}-\frac{\delta}{4}(1+\varepsilon) &\frac{\delta}{2}(1+\varepsilon)
\end{array}
 \right].
\end{equation*}

Similarly, we obtain the backward recursion probability transition  matrix as
\renewcommand{\arraystretch}{1.2}
\begin{equation*}
\boldsymbol{M}_\beta=\left[
\begin{array}{c c c}
 \frac{1}{2}\bar{\varepsilon} & \frac{1}{2}\bar{\varepsilon} &  \varepsilon \\
\frac{1}{2}\bar{\varepsilon} & \frac{1}{2}\bar{\varepsilon} &  \varepsilon \\
\frac{1}{4}\bar{\varepsilon}\bar{\delta} & \frac{1}{4}\bar{\varepsilon}\bar{\delta} &\varepsilon+\frac{1}{2}\delta-\frac{1}{2}\delta\varepsilon
\end{array}
 \right].
\end{equation*}
\begin{table}[!t]
\caption{Transfer functions of ISI channels}
\vspace{-3.5ex}
\begin{center}
\begin{tabular}{cc}
	\toprule
CH-\RomanNumeralCaps{1} & $g(\delta,\varepsilon)=\frac{4\varepsilon^2}{(\delta\varepsilon - \delta + 2)^2}$\\[0.2cm]
\hline  
CH-\RomanNumeralCaps{2}& $ g(\delta,\varepsilon)=\frac{4\varepsilon^3(4\delta\varepsilon - 4\delta - \delta^2\varepsilon + \delta^2 + 4)}{(\delta^2\varepsilon^3 - \delta^2\varepsilon^2 + 2\delta\varepsilon^2 - 2\delta + 4)^2}$ \\[0.2cm]
\hline 

   CH-\RomanNumeralCaps{3} & $g(\delta,\varepsilon)$ see top of the page \\[0.05cm]
\bottomrule
\end{tabular}
\end{center}
\label{Tab:erasureproba}
\vspace{-3.5ex}
\end{table}
Now, denote the steady state distribution vector of the forward and backward Markov chain by $\boldsymbol{\pi}_{a}$, which can be computed as the solution to
\begin{equation}
\boldsymbol{\pi}_{a}=\boldsymbol{M}_{a}\cdot \boldsymbol{\pi}_{a}\,,
\end{equation}
where $a\in\{\alpha,\beta\}$. Also, define the $|\mathcal{M}_\alpha|\times|\mathcal{M}_\beta|$ matrix  $\boldsymbol{T}$ with entries
$$
T_{ij}=p(\Xtilde=\que| \alpha_t=\boldsymbol{m}_{\alpha}^{(i)},\beta_{t+1}=\boldsymbol{m}_{\beta}^{(j)})\,,
$$
where $\Xtilde$ is the message passed from the channel detector to the decoder. In words, $T_{ij}$ is the average (extrinsic)  probability that the symbol passed by the channel detector to the decoder is an erasure  given $\alpha_t$ and $\beta_{t+1}$. Then, the extrinsic erasure probability of a message from the channel detector to the decoder is given by 
$$
g(\delta,\varepsilon)=\boldsymbol{\pi}_\alpha \cdot \boldsymbol{T}\cdot \boldsymbol{\pi}_\beta\,,
$$
which we refer to as the input/output transfer function of the channel detector. 

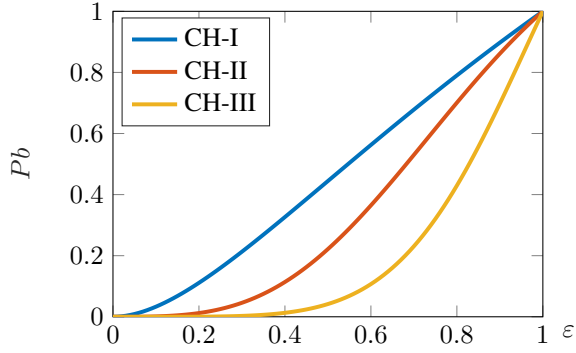
\begin{figure}[t]
\centering
%
%
\definecolor{mycolor1}{rgb}{0.00000,0.44700,0.74100}%
\definecolor{mycolor2}{rgb}{0.85000,0.32500,0.09800}%
\definecolor{mycolor3}{rgb}{0.92900,0.69400,0.12500}%
\begin{tikzpicture}
\begin{axis}[%
width=2.25in,
height=1.6in, 
at={(0.2in,0.271in)},
scale only axis,
xmin=0,
xmax=1,
xlabel style={font=\color{white!15!black}},
every axis x label/.style={
    at={(ticklabel* cs:1.06)},
    anchor=north,},
xlabel={$\varepsilon$},
ymin=0,
ymax=1,
ylabel style={font=\color{white!15!black}},
ylabel={$Pb$},
axis background/.style={fill=white},
legend style={at={(axis cs: 0.02,0.98)}, anchor=north west, legend cell align=left, align=left, draw=white!15!black}
]
\addplot [color=mycolor1, line width=1.5pt]
  table[row sep=crcr]{%
0	0\\
0.01	0.000392118419762768\\
0.02	0.00153787004998078\\
0.03	0.00339334527288152\\
0.04	0.00591715976331361\\
0.05	0.0090702947845805\\
0.06	0.0128159487362051\\
0.07	0.0171193990741549\\
0.08	0.0219478737997256\\
0.09	0.0272704317818365\\
0.1	0.0330578512396694\\
0.11	0.0392825257690123\\
0.12	0.0459183673469388\\
0.13	0.0529407157960686\\
0.14	0.0603262542320714\\
0.15	0.0680529300567108\\
0.16	0.0760998810939358\\
0.17	0.0844473664986485\\
0.18	0.0930767020970985\\
0.19	0.101970199844644\\
0.2	0.111111111111111\\
0.21	0.120483573526398\\
0.22	0.130072561139479\\
0.23	0.139863837662767\\
0.24	0.149843912591051\\
0.25	0.16\\
0.26	0.170319979843789\\
0.27	0.180792361584723\\
0.28	0.19140625\\
0.29	0.202151313022054\\
0.3	0.21301775147929\\
0.31	0.223996270613601\\
0.32	0.235078053259871\\
0.33	0.246254734580813\\
0.34	0.257518378257964\\
0.35	0.268861454046639\\
0.36	0.280276816608997\\
0.37	0.291757685545314\\
0.38	0.303297626549044\\
0.39	0.314890533616272\\
0.4	0.326530612244898\\
0.41	0.338212363563201\\
0.42	0.349930569331482\\
0.43	0.361680277764194\\
0.44	0.373456790123457\\
0.45	0.38525564803805\\
0.46	0.397072621504973\\
0.47	0.408903697533435\\
0.48	0.420745069393718\\
0.49	0.432593126435746\\
0.5	0.444444444444444\\
0.51	0.456295776501031\\
0.52	0.46814404432133\\
0.53	0.479986330044\\
0.54	0.491819868443245\\
0.55	0.503642039542144\\
0.56	0.515450361604208\\
0.57	0.527242484482129\\
0.58	0.539016183303958\\
0.59	0.550769352478146\\
0.6	0.5625\\
0.61	0.574206242043131\\
0.62	0.585886297820454\\
0.63	0.597538484700215\\
0.64	0.609161213563355\\
0.65	0.620752984389348\\
0.66	0.632312382058354\\
0.67	0.643838072358278\\
0.68	0.655328798185941\\
0.69	0.666783375932215\\
0.7	0.678200692041522\\
0.71	0.689579699736671\\
0.72	0.700919415900487\\
0.73	0.712218918106185\\
0.74	0.723477341788876\\
0.75	0.73469387755102\\
0.76	0.745867768595041\\
0.77	0.756998308276677\\
0.78	0.768084837773009\\
0.79	0.77912674385943\\
0.8	0.790123456790123\\
0.81	0.801074448276915\\
0.82	0.811979229561647\\
0.83	0.822837349577473\\
0.84	0.833648393194707\\
0.85	0.844411979547115\\
0.86	0.855127760434732\\
0.87	0.865795418799508\\
0.88	0.876414667270258\\
0.89	0.886985246773607\\
0.9	0.897506925207756\\
0.91	0.907979496176092\\
0.92	0.918402777777778\\
0.93	0.928776611452656\\
0.94	0.939100860877883\\
0.95	0.949375410913872\\
0.96	0.959600166597251\\
0.97	0.969775052178618\\
0.98	0.979900010203041\\
0.99	0.989975000631297\\
1	1\\
};
\addlegendentry{CH-I}
\addplot [color=mycolor2, line width=1.5pt]
  table[row sep=crcr]{%
0	0\\
0.01	1.02989597787974e-06\\
0.02	8.47654121842314e-06\\
0.03	2.94027373458242e-05\\
0.04	7.15608731708898e-05\\
0.05	0.00014337339784757\\
0.06	0.000253910154499653\\
0.07	0.000412862524484586\\
0.08	0.000630514353962322\\
0.09	0.000917709657284172\\
0.1	0.0012858171157589\\
0.11	0.00174669141541448\\
0.12	0.00231263149325404\\
0.13	0.00299633578798051\\
0.14	0.0038108546179926\\
0.15	0.00476953983637273\\
0.16	0.00588599193931698\\
0.17	0.00717400483070762\\
0.18	0.00864750847100046\\
0.19	0.0103205096629881\\
0.2	0.01220703125\\
0.21	0.0143210500234133\\
0.22	0.0166764336556752\\
0.23	0.019286876992112\\
0.24	0.0221658380493461\\
0.25	0.0253264740799367\\
0.26	0.0287815780716856\\
0.27	0.0325435160557291\\
0.28	0.0366241655999312\\
0.29	0.0410348558630941\\
0.3	0.045786309581053\\
0.31	0.0508885873477931\\
0.32	0.0563510345433603\\
0.33	0.0621822312455862\\
0.34	0.0683899454446402\\
0.35	0.0749810898583121\\
0.36	0.0819616826219203\\
0.37	0.0893368121000745\\
0.38	0.09711060603847\\
0.39	0.105286205242777\\
0.4	0.113865741938823\\
0.41	0.122850322934054\\
0.42	0.132240017664996\\
0.43	0.142033851179631\\
0.44	0.152229802067513\\
0.45	0.162824805314555\\
0.46	0.173814760024093\\
0.47	0.185194541911419\\
0.48	0.196958020445851\\
0.49	0.209098080482887\\
0.5	0.221606648199446\\
0.51	0.23447472111777\\
0.52	0.247692401978635\\
0.53	0.26124893620221\\
0.54	0.275132752655406\\
0.55	0.289331507427977\\
0.56	0.303832130306085\\
0.57	0.31862087362156\\
0.58	0.333683363147632\\
0.59	0.349004650707538\\
0.6	0.364569268160951\\
0.61	0.380361282434644\\
0.62	0.396364351267916\\
0.63	0.41256177935008\\
0.64	0.428936574536432\\
0.65	0.445471503840428\\
0.66	0.462149148913115\\
0.67	0.478951960735885\\
0.68	0.495862313269197\\
0.69	0.51286255581774\\
0.7	0.529935063891365\\
0.71	0.547062288360785\\
0.72	0.564226802727266\\
0.73	0.581411348346069\\
0.74	0.598598877464127\\
0.75	0.615772593953017\\
0.76	0.632915991638648\\
0.77	0.65001289014901\\
0.78	0.667047468220639\\
0.79	0.684004294423085\\
0.8	0.700868355278415\\
0.81	0.717625080769618\\
0.82	0.734260367247573\\
0.83	0.750760597760958\\
0.84	0.767112659847028\\
0.85	0.783303960833601\\
0.86	0.799322440713771\\
0.87	0.815156582664887\\
0.88	0.830795421292118\\
0.89	0.846228548684584\\
0.9	0.861446118378499\\
0.91	0.876438847327183\\
0.92	0.891198015982068\\
0.93	0.905715466592194\\
0.94	0.919983599831982\\
0.95	0.933995369868577\\
0.96	0.947744277980634\\
0.97	0.961224364840317\\
0.98	0.97443020156941\\
0.99	0.987356879678985\\
1	1\\
};
\addlegendentry{CH-II}

\addplot [color=mycolor3, line width=1.5pt]
  table[row sep=crcr]{%
0	0\\
0.01	1.00505612397754e-10\\
0.02	3.23271674324059e-09\\
0.03	2.46767175086555e-08\\
0.04	1.04539307131698e-07\\
0.05	3.20746801956044e-07\\
0.06	8.02480168616834e-07\\
0.07	1.74407858712692e-06\\
0.08	3.41943661929531e-06\\
0.09	6.19691903812621e-06\\
0.1	1.05548160153264e-05\\
0.11	1.70973597221737e-05\\
0.12	2.65713214269624e-05\\
0.13	3.98832058211287e-05\\
0.14	5.81170565250905e-05\\
0.15	8.25528834611556e-05\\
0.16	0.000114685718980349\\
0.17	0.000156245305237265\\
0.18	0.000209216410265737\\
0.19	0.000275859764461602\\
0.2	0.000358733602670545\\
0.21	0.000460715789753312\\
0.22	0.000585026499303446\\
0.23	0.000735251406072682\\
0.24	0.000915365342568403\\
0.25	0.00112975635918326\\
0.26	0.0013832501150626\\
0.27	0.00168113451368188\\
0.28	0.0020291844827751\\
0.29	0.00243368678281849\\
0.3	0.00290146471173827\\
0.31	0.00343990255589925\\
0.32	0.00405696961878375\\
0.33	0.00476124363915022\\
0.34	0.00556193338995468\\
0.35	0.00646890022803879\\
0.36	0.0074926783426802\\
0.37	0.00864449342874007\\
0.38	0.00993627948754072\\
0.39	0.0113806934360158\\
0.4	0.0129911271823857\\
0.41	0.0147817168049603\\
0.42	0.0167673484500343\\
0.43	0.0189636605456499\\
0.44	0.0213870419107229\\
0.45	0.0240546253241808\\
0.46	0.0269842761069088\\
0.47	0.0301945752610382\\
0.48	0.0337047967070808\\
0.49	0.0375348781602841\\
0.5	0.0417053851940351\\
0.51	0.0462374680508838\\
0.52	0.0511528107814889\\
0.53	0.0564735723191936\\
0.54	0.0622223191336689\\
0.55	0.0684219491517332\\
0.56	0.0750956066875909\\
0.57	0.0822665881887794\\
0.58	0.0899582386784029\\
0.59	0.0981938388589377\\
0.6	0.106996482938044\\
0.61	0.116388947342228\\
0.62	0.126393550599486\\
0.63	0.137032004796565\\
0.64	0.148325259149347\\
0.65	0.160293336364852\\
0.66	0.172955162619061\\
0.67	0.186328392124414\\
0.68	0.200429227412281\\
0.69	0.215272236606742\\
0.7	0.230870169113892\\
0.71	0.247233771292815\\
0.72	0.26437160380727\\
0.73	0.282289862477767\\
0.74	0.300992204558815\\
0.75	0.32047958245232\\
0.76	0.340750086932128\\
0.77	0.361798801993464\\
0.78	0.383617673451532\\
0.79	0.406195393393344\\
0.8	0.429517302533699\\
0.81	0.45356531243865\\
0.82	0.478317849456661\\
0.83	0.503749822038749\\
0.84	0.529832612934632\\
0.85	0.556534097523554\\
0.86	0.583818689278072\\
0.87	0.611647413069686\\
0.88	0.6399780067104\\
0.89	0.668765050788794\\
0.9	0.697960126508051\\
0.91	0.727512000872578\\
0.92	0.757366838205657\\
0.93	0.787468436619604\\
0.94	0.817758487709181\\
0.95	0.84817685740519\\
0.96	0.878661885615188\\
0.97	0.909150701998268\\
0.98	0.939579554976874\\
0.99	0.969884150885675\\
1	1\\
};
\addlegendentry{CH-III}
\end{axis}
\end{tikzpicture}%
\caption{Transfer functions for the three channels for $\delta=1$.}
\label{transferFuncfig}
\vspace{-5mm}
\end{figure}

In Table~\ref{Tab:erasureproba}, we give the transfer function of three ISI channels with impulse response given in Table~\ref{Tab:channels}, of memory $\nu=1$ (CH-I), $2$ (CH-II), and $4$ (CH-III), respectively, with erasures. The transfer functions are  depicted in  Fig.~\ref{transferFuncfig} for $\delta=1$.

%

\section{BP and MAP Thresholds for SC-LDPC Codes over ISI Channels}

The receiver applies joint iterative message passing detection and decoding between the channel and the decoder, in terms of log-likelihood ratios ($L$-values) as illustrated in Fig.~\ref{messages1}. At iteration $\ell$ and spatial position $t$, first the messages $L^{(\ell)}_{\mathcal{H} \rightarrow
\text{v},t}$ from the channel detector to the variable nodes are updated with the BCJR algorithm, using the received sequence $\bs{y}$ and the incoming messages $L^{(\ell-1)}_{ \text{v} \rightarrow \mathcal{H},t}$ from the previous iteration of the channel decoder, where $L^{(0)}_{ \text{v} \rightarrow \mathcal{H},t}=0$. Then $I_\text{c}$ decoding iterations are performed between the variable nodes and check nodes of the SC-LDPC code. At the variable node updates in decoding iteration $i$, the messages $L^{(\ell)}_{\mathcal{H} \rightarrow
\text{v},t}$ from the channel are combined with the incoming messages from the check nodes  $L^{(i-1)}_{\text{c} \rightarrow
\text{v},t}$ to produce the outgoing messages $L^{(i)}_{\text{v} \rightarrow \text{c},t}$. These are then used at the check node updates to produce the messages $L^{(i)}_{\text{c} \rightarrow
\text{v},t}$. Using $I_\text{c}>1$ decoding iterations between the channel detector updates reduces the overall complexity of the receiver.
\subsection{Density Evolution for ISI Channels with Erasures}

In case of ISI channels with erasures, all $L$-values exchanged in the message passing receiver can take the values $L=-\infty$ or $L=+\infty$ if the corresponding variable is a known $-1$ or $+1$, respectively, or $L=0$ if the variable is an erasure. For this reason, the messages $-1$, $+1$ and '$?$' can be exchanged instead of $L$-values, and density evolution is equivalent to tracking the evolution of the erasure probabilities of the variables.

The probability $q^{(\ell)}_{\mathcal{H} ,t}$ that an outgoing message $L^{(\ell)}_{\mathcal{H} \rightarrow \text{v},t}$ of the channel detector is an erasure is given by the transfer function derived in Section~\ref{sec:transferfunction}, i.e.,
$$
q^{(\ell)}_{\mathcal{H},t} = g \left(\delta_t^{(\ell-1)},\varepsilon \right) \ ,
$$
where $\delta_t^{(\ell-1)}$ denotes the erasure probability of the incoming messages $L^{(\ell-1)}_{\mathcal{H} \rightarrow
\text{v},t}$ from the previous code-channel iteration.

For the decoding of the SC-LDPC code, the density evolution equations can be derived follwing the approach in \cite{5205688}.
At a variable node update, the average probability that a message $L^{(i)}_{\text{v} \rightarrow \text{c},t}$ is an erasure is given by
\begin{equation}\label{eq:DE1}
p^{(i)}_{t} = q^{(\ell)}_{\mathcal{H},t} \cdot \frac{1}{m+1} \sum_{j=0}^{m} \left( q^{(i-1)}_{\text{c},t+j} \right)^{d_v-1} \ ,
\end{equation}
where 
\begin{equation}\label{eq:DE2}
q^{(i)}_{\text{c},t} = \frac{1}{m+1} \sum_{j=0}^{m} \left( 1 - \left(1 - p^{(i)}_{t-j}\right)^{d_c-1} \right) 
\end{equation}
denotes the average erasure probability of the messages $L^{(i)}_{\text{c} \rightarrow \text{v},t}$ computed at a check node update. 
Equations \eqref{eq:DE1} and \eqref{eq:DE2} are valid for a $(d_v,d_c)$-regular LDPC code that is spatially coupled in such a way that edges are spread uniformly over $m+1$ spatial positions.

A message $L^{(\ell)}_{ \text{v} \rightarrow \mathcal{H},t}$ passed to the channel detector after $I_\text{c}$ decoding iterations is erased if all $d_v$ messages  $L^{(I_\text{c})}_{\text{c} \rightarrow \text{v},t}$ are erased. The corresponding erasure probability is hence equal to
\[
\delta^{(\ell)}_{t} =  \frac{1}{m+1} \sum_{j=0}^{m} \left( q^{(i-1)}_{\text{c},t+j} \right)^{d_v} \ .
\]
 The average GEXIT function can be expressed   as \cite{Wan2008}
 $$
 \frac{1}{n}\frac{d H(\mathcal{X}|\mathcal{Y},S_0)}{d\varepsilon}=\frac{1}{n}\sum_{i=1}^{n}H(Z_i|\mathcal{Y}_{\sim i},S_0) \ ,
 $$
 where the channel parameter is taken to be $\text{h}_i=\varepsilon$.   
 
 From the GEXIT function we can thus define the generalized BP EXIT (GB-EXIT) function, $h_i^{\text{BP}}$, as the joint iterative decoding of $Z_i$ from $\mathcal{Y}_{\sim i}$ and $S_0$. The exact GB-EXIT function is computed in the same fashion as in \cite{Wan2008}. That is,
 $$
   \text{h}^{\text{BP}}= \sum_{\mathclap{i,j}}P(\boldsymbol{m}_\alpha^{(i)})H(Z_i|\alpha_t=\boldsymbol{m}_\alpha^{(i)},\beta_{t+1}=\boldsymbol{m}_\beta^{(j)})P(\boldsymbol{m}_\beta^{(j)})  \ .
 $$
The GB-EXIT function is used  as described in \cite{Wan2008} to compute an upper bound $\varepsilon^{\text{MAP}}$ on the MAP threshold.
 The SIR also can be computed using the state distribution of the Markov chain from Section~\ref{sec:transferfunction}. Using the notation $\mathcal{X}$ for the sequence $X_1X_2\dots X_t$ the SIR is computed using the expression
$I(\mathcal{X;Y})=H(\mathcal{Y})-H(\mathcal{Y|X})$, with $H(\mathcal{Y|X})=h_b(\varepsilon)$, where $h_b(\cdot)$ denotes the binary entropy function. The entropy rate $H(\mathcal{Y})$ can be computed by using the definition \cite{10.5555/959778}
$$H(\mathcal{Y})=\lim_{t \to \infty} H(Y_t|Y^{t-1})=\sum_{\mathclap{i=1}}^{|\mathcal{M}_\alpha|}P_{\alpha_t}(\boldsymbol{m}_\alpha^{(i)})H(Y|\alpha_t=\boldsymbol{m}_\alpha^{(i)}).
$$
$P_{\alpha_t}(\boldsymbol{m}_\alpha^{(i)})$ is the $i$-th entry in the steady state distribution $\boldsymbol{\pi}_{\alpha1}$ computed without using \textit{a-priori} inputs from the code. This can be obtained from the steady state distribution $\boldsymbol{\pi}_{\alpha}(\varepsilon,\delta)$ by setting the erasure probability from the code to $1$, i.e., $\boldsymbol{\pi}_{\alpha1}(\varepsilon)=\boldsymbol{\pi}_{\alpha}(\varepsilon,1)$.
 
\begin{table}[!t]
\caption{Thresholds for the $(3,6)$ code for different ISI channels}
\vspace{-4ex}
\begin{center}
\begin{tabular}{m{1.21cm}|m{0.69cm}m{0.69cm}|m{0.69cm}m{0.69cm}|m{0.69cm}m{0.69cm}}
\toprule
{}&\multicolumn{2}{c|}{CH-\RomanNumeralCaps{1}}&\multicolumn{2}{c|}{CH-\RomanNumeralCaps{2}}&\multicolumn{2}{c}{CH-\RomanNumeralCaps{3}}\\
  {}   & Erasures  & AWGN & Erasures & AWGN & Erasures & AWGN\\
  \hline
  $\varepsilon^{\text{BP}},\gamma^{\text{BP}}$   & $0.5689$ &$1.703$& $0.7055$& $ 2.598$ &$0.8254$&$5.474$\\
  $\varepsilon^{\text{MAP}},\gamma^{\text{MAP}}$ & $0.6387$&$1.1600$ & $0.7519$&$ 1.5090$&$0.8482$&$2.9750$\\
  $\text{h}^{\text{BP}}$      & $0.8530$&$0.8510$  & $1.5870$&  $ 1.5147$ &$3.3010$&$2.9177$ \\
$\text{h}^{\text{max}}$      & $1.5000$& $1.5000$ & $2.2500$& $2.2500$ &$4.0000$&$4.0000$ \\
$\text{h}^{\text{BP}}/\text{h}^{\text{max}}$ & $0.5689$ &$0.5673$& $0.7055$& $ 0.6732$&$0.8254$&$0.7294$\\
$\text{h}^{\text{MAP}}$      & $0.9580$&$0.9195$  &  $1.6918$& $ 1.6200$&$3.3926$&$3.2146$\\
$\varepsilon^{\text{SIR}},\gamma^{\text{SIR}}$   & $0.6404$ &$0.8230 $  & $0.7530$&$ 1.4370$ &$0.8506$&$2.9600 $\\
\bottomrule
\end{tabular}
\end{center}
\label{compare36}
\vspace{-5ex}
\end{table}
 
\subsection{Density Evolution for ISI Channels with AWGN}

For channels with AWGN, no explicit transfer functions are available for computing the probability densities $p(L^{(\ell)}_{\mathcal{H} \rightarrow \text{v},t})$ of the outgoing messages at the channel detector. For this reason we use Monte Carlo simulations for the BCJR detector to determine these densities from the incoming densities $p(L^{(\ell)}_{ \text{v} \rightarrow \mathcal{H},t})$ and the noise distribution.

For the message passing decoding of the SC-LDPC code we use discretized density evolution \cite{Sae2001}, which is exact up to the numerical accuracy and the resolution of the underlying quantization. 
During the $i$-th iteration within the code, at the variable nodes the message densities $p(L^{(i)}_{\text{v} \rightarrow \text{c},t})$ along an edge can be obtained as the convolution of the density $p(L^{(\ell)}_{\mathcal{H} \rightarrow
\text{v},t})$ from the channel with the $d_v-1$ densities $p(L^{(i)}_{\text{c} \rightarrow \text{v},t})$ of incoming messages from the other edges. This can be implemented efficiently using the FFT. 
At the check nodes, the densities $p(L^{(i)}_{\text{c} \rightarrow \text{v},t})$ are computed from the $d_c-1$ incoming densities $p(L^{(i)}_{\text{v} \rightarrow \text{c},t})$ in a nested fashion from a two-dimensional lookup table \cite{Sae2001}.
%
%
%
%
 After the $I_\text{c}$ iterations between check and variable nodes are completed, the  density $p(L^{(\ell)}_{ \text{v} \rightarrow \mathcal{H},t})$ of messages passed to the channel is obtained as the convolution of all $d_v$ message densities $p(L^{(I_\text{c})}_{\text{c} \rightarrow \text{v},t})$.
\begin{table*}[t]

\caption{Thresholds of regular codes with spatial coupling for ISI channels with erasures and AWGN}
\vspace{-3.5ex}
\begin{center}
\begin{tabular}{cc|cccccc|cccccc}
\toprule
{}&{}&\multicolumn{6}{c|}{Erasures}&\multicolumn{6}{c}{AWGN}\\

Code&Channel&$\varepsilon^{\text{BP}}$&$\varepsilon^{1}$&$\varepsilon^{3}$&$\varepsilon^{6}$&$\varepsilon^{\text{MAP}}$&$\varepsilon^{\text{SIR}}$&$\gamma^{\text{BP}}$&$\gamma^{1}$&$\gamma^{3}$&$\gamma^{10}$&$\gamma^{\text{MAP}}$&$\gamma^{\text{SIR}}$\\
\otoprule
\multirow{3}*{$(3,6)$}&CH-\RomanNumeralCaps{1}&$0.5689$ &$0.6386$ &$0.6386$&$0.6386$ &$0.6387$&$0.6404$&$1.703$ &$1.330$ &$1.240$&$1.178$&$1.160$&$0.823$\\
                      &CH-\RomanNumeralCaps{2}&$0.7055$ &$0.7519$ &$0.7519$&$0.7519$ &$0.7519$&$0.7530$&$2.598$ &$1.598$ &$1.587$&$1.535$&$1.509$&$1.437$\\
                       &CH-\RomanNumeralCaps{3}&$0.8254$&$0.8482$&$0.8482$&$0.8482$&$0.8482$&$0.8506$&$5.474$ &$3.019$&$3.010$&$2.998$&$2.975$&$2.960$\\
\hline                      
\multirow{3}*{$(4,8)$}&CH-\RomanNumeralCaps{1}&$0.5100$& $0.6399$&$0.6401$&$0.6401$&$0.6404$&$0.6404$&$2.441$ &$0.896$&$0.877$ &$0.866$&$0.853$&$0.823$\\
                      &CH-\RomanNumeralCaps{2}&$0.6618$&$0.7528$&$0.7528$&$0.7528$&$0.7530$&$0.7530$&$3.596$ &$1.494$ &$1.478$&$1.458$&$ 1.448$&$1.437$\\
            &CH-\RomanNumeralCaps{3}&$0.7997$&$0.8501$&$0.8501$&$0.8501$&$0.8501$&$0.8506$&$6.745$ &$3.100$&$3.061$&$2.993$&$2.963$&$2.960$\\
\hline       
\multirow{3}*{$(5,10)$}&CH-\RomanNumeralCaps{1}&$0.4647$& $0.6400$& $0.6403$& $0.6403$& $0.6404$& $0.6404$&$3.029$ &$0.877$ &$0.852$ &$0.844$&$0.834$&$0.823$\\
                      &CH-\RomanNumeralCaps{2}&$0.6275$& $0.7526$& $0.7529$& $0.7529$& $0.7530$& $0.7530$&$4.348$ &$1.483$&$1.461$ &$1.450$&$1.437$&$1.437$\\
                       &CH-\RomanNumeralCaps{3}&$0.7775$&$0.8503$&$0.8503$&$0.8503$&$0.8503$&$0.8506$&$7.550$ &$3.036$ &$3.000$&$2.987$&$2.960$&$2.960$\\
\hline  
\multirow{3}*{$(6,12)$}&CH-\RomanNumeralCaps{1}&$0.4647$& $0.6378$&$0.6403$&$0.6403$&$0.6404$&$0.6404$&$3.517$ &$0.853$ &$0.844$ &$0.829$&$0.823$&$0.823$\\
                      &CH-\RomanNumeralCaps{2}& $0.6000$& $0.7504$& $0.7529$& $0.7530$& $0.7530$& $0.7530$&$4.938$ &$1.483$&$1.461$ &$1.450$&$1.437$&$1.437$\\
                       &CH-\RomanNumeralCaps{3}&$0.7588$&$0.8504$&$0.8504$&$0.8504$&$0.8504$&$0.8506$&$8.123$ &$3.036$ &$2.990$&$2.980$&$2.960$&$2.960$\\
\bottomrule                      
\end{tabular} 
\end{center}
\label{Tab:SpatialCombined} 
\vspace{-5ex}
\end{table*}

The GEXIT function for the BIAWGNC with ISI can be defined as \cite{Ngu2012}
$$
G(\text{h})=\frac{1}{n}\frac{d H(\mathcal{X}|\mathcal{Y},S_0)}{d\text{h}}
$$
where
\begin{equation}\label{entropyGauss}
\begin{aligned}
\text{h}&=H(Z_i|Y_i)=H(Z_i)-I(Y_i;Z_i),
\end{aligned}
\end{equation}
%
The entropy $\text{h}$ is a function of the channel parameter $\varepsilon$ which is chosen for convenience to be $\varepsilon=-\frac{1}{2\sigma^2}$~\cite{Ngu2012}.
The signal-to-noise ratio parameter used in this paper, $\gamma$ is the ratio of information bit energy, $E_b$, to the noise spectrum density, $N_o$, in decibels. Using the GBP-EXIT function, an upper bound $\gamma^{\text{MAP}}$ on the MAP threshold is computed  as described in \cite{Ngu2012}. The SIR is computed numerically using the method described in \cite{Arn2006}.




\section{Numerical Results}
 If we compare the BP thresholds of the channels by looking at the channel erasure probability $\varepsilon$ in Table \ref{compare36} for the regular (3,6) code we observe that CH-\RomanNumeralCaps{3} has the best performance (the threshold occurs at the highest erasure probability) followed by CH-{\RomanNumeralCaps{2}} while CH-{\RomanNumeralCaps{1}} has the worst performance. On the other hand, if we look at the Gaussian channel we notice that the ordering is reversed with CH-{\RomanNumeralCaps{1}} having the best threshold (at lowest SNR) and CH-{\RomanNumeralCaps{3}} has the worst performance. Thus we see an apparent inconsistency in the performance of the channels when erasures are changed to AWGN. This is also true for the MAP threshold and the SIR.
 
 In Table \ref{Tab:SpatialCombined} we see the same phenomenon when different codes are used. That is, the channel ranking for the BP and MAP thresholds as well as the SIR is reversed for all codes when changing from erasures to AWGN. 

But we can also characterize the thresholds in terms of the entropy $\text{h}=H(Z_i|Y_i)$. This is defined in \eqref{entropyGauss} for the Gaussian channel, while for the erasure channel we observe that 
$$
H(Z_i|Y_i)=\bar{\varepsilon}H(Z_i|Y_i\neq \que)+\varepsilon H(Z_i|Y_i=\que)=\varepsilon H(Z_i),
$$
since $Z_i$ is known with certainty when $Y_i$ is not erased and it is independent of it when it is erased. 

Representing the thresholds in terms of entropy, we now observe in Table \ref{compare36} that the ranking of the channels is unchanged when changing between erasures and AWGN. I.e., CH-{\RomanNumeralCaps{3}} has the best performance with both AWGN and erasures by having thresholds which are at the highest entropy, while CH-{\RomanNumeralCaps{1}} has the worst performance. This could be attributed to the fact the CH-{\RomanNumeralCaps{3}} has the highest $H(Z_i)$ (labelled $\text{h}^{\text{max}}$ in the table). But if we normalize the threshold by dividing by $\text{h}^\text{max}$ the ranking is unchanged. It is interesting to note that the BP entropies $\text{h}^{\text{BP}}$ for erasures and AWGN are relatively close to each other for a given ISI channel. On the other hand, their gap is still too large for making an accurate prediction from the erasure to the Gaussian case.

In Table \ref{Tab:SpatialCombined} we can also see that with uncoupled regular codes the MAP threshold improves with increasing variable node degree, while the BP threshold becomes worse. With spatially-coupled codes, we observe that the BP thresholds approach the MAP thresholds of the uncoupled codes for all three channels. It is also interesting to see that for the (5,10) and (6,12) code the MAP threshold is equal or very close to the SIR for all three channels. This demonstrates that with spatial coupling we can universally approach the SIR of different ISI channels using a single code. This makes spatially coupled codes superior to uncoupled irregular codes that need to be optimized for a particular ISI channel, which cannot guarantee robust performance if the channel is changed.     

\section{Conclusions}
We have derived exact transfer function for three different channels using a method which can be applied to any arbitrary ISI channel.  We have further shown that to compare the behaviour of ISI channels with erasures and AWGN, the proper parameter is the entropy $H(Z_i|Y_i)$, by which we can observe a consistent behaviour. Finally we have shown numerically that with spatially coupled LDPC codes we can universally achieve the SIR of different ISI channels.  



\end{document}